\documentclass[preprint]{aastex62} 
\usepackage{graphicx,natbib,color,amsmath,subfigure} 
\usepackage{ulem} 
\newcommand{\blank}[1]{} 
\usepackage{color} 
\usepackage{hyperref} 
\usepackage[autostyle]{csquotes} 
 
\received{March 27, 2019} 
\revised{April 30, 2019} 
\accepted{May 13, 2019} 
\begin{document} 
\title{ALMA detection of dark chromospheric holes in the quiet Sun} 
%\correspondingauthor{Maria Loukitcheva} 
%\email{lukicheva@mps.mpg.de, maria.lukicheva@spbu.ru} 

\author[0000-0002-0786-7307]{Maria A. Loukitcheva} 
\affiliation{Max-Planck-Institut f\"ur Sonnensystemforschung, Justus-von-Liebig-Weg 3, 37077 G\"ottingen, Germany} 
\affiliation{Saint Petersburg branch of Special Astrophysical Observatory, Pulkovskoye chaussee 65/1, St. Petersburg 196140, Russia} 
\affiliation{Saint Petersburg State University, 7/9 Universitetskaya nab., St. Petersburg 199034, Russia} 
\author{Stephen M. White} 
\affiliation{Space Vehicles Directorate, Air Force Research Laboratory, Albuquerque, NM, USA} 
\author{Sami K. Solanki} 
\affiliation{Max-Planck-Institut f\"ur Sonnensystemforschung, Justus-von-Liebig-Weg 3, 37077 G\"ottingen, German} 
\affiliation{School of Space Research, Kyung Hee University, Yongin, Gyeonggi 446-701, Korea}

\begin{abstract} 
We present Atacama Large Millimeter/submillimeter Array (ALMA) observations of a quiet-Sun region at a wavelength of 3~mm, obtained during the first solar ALMA cycle on April 27, 2017, and compare them with available chromospheric observations in the UV and visible as well as with photospheric magnetograms. ALMA images clearly reveal the presence of distinct particularly dark/cool areas in the millimeter maps having temperatures as low as 60\% of the normal quiet Sun at 3~mm, which are not seen in the other data.  We speculate that ALMA is sensing cool chromospheric gas, whose presence had earlier been inferred from infrared CO spectra. 
\end{abstract} 
 
 
\section{Introduction} 
 
Observations of the quiet Sun at submillimeter (submm) and millimeter (mm) wavelengths can provide essential diagnostics of the physical conditions in the solar chromosphere \citep[see, e.g., ][and references therein]{2018A&A...620A.124D, 2004A&A...419..747L}. Submm and mm continua originate from the low to mid chromosphere and allow a rather straightforward measurement of the gas temperature at these heights, as the radiation at these wavelengths is coupled linearly to the electron temperature owing to its formation in local thermodynamic equlibrium (LTE) and in the Rayleigh-Jeans limit.  
 
Prior to the advent of the Atacama Large Millimeter/submilimeter Array \citep[ALMA, ][]{2009IEEEP..97.1463W} quiet-Sun observations at these wavelengths were extremely rare due to the poor (from tens of arcseconds to arcminutes) spatial resolution of most instruments operating at submm/mm wavelengths. Unique observations with the 10-element Berkeley-Illinois-Maryland Array (BIMA) at 3.5 mm with a resolution of around 10\arcsec, reported  in \citet{2006A&A...456..697W} and \citet{2009A&A...497..273L}, resulted in the first successful interferometric mapping of chromospheric structure at mm wavelengths. Strong morphological similarities between mm brightness and chromospheric emissions in the Ca II K-line, the far-ultraviolet continuum, and the photospheric magnetic field were found on these spatial scales, at which the chromospheric network can just about be distinguished from the internetwork. However, the spatial resolution and scale coverage of the BIMA interferometer was insufficient to clearly resolve the arcsecond-scale quiet-Sun fine structure and reliably study correlations with the chromospheric features seen in other spectral domains. 
 
In this letter we present results of ALMA observations of the quiet Sun in conjunction with co-temporal images at UV, EUV, and visible wavelengths from the Solar Dynamics Observatory \citep[SDO, ][]{2012SoPh..275....3P}, the Interface Region Imaging Spectrograph \citep[IRIS, ][]{2014SoPh..289.2733D}, and the instruments of the Global Oscillation Network Group (GONG), including the Big Bear Solar Observatory (BBSO) and the Cerro Tololo Interamerican Observatory (CTIO) in Chile. Observational data are summarized in Sect.~\ref{obs}. In Section~\ref{res} we present the results of comparison of the mm images with the data from other spectral domains and report the first detection of particular dark (i.e. cool) regions in ALMA images. Results are discussed and conclusions are drawn in Sect.~\ref{discussion}. 
 
\section{Data Collection and Reduction}\label{obs} 
%\subsection{ALMA} 
 
The ALMA interferometric data were acquired on 2017 April 27 in Band 3 (100 GHz, i.e., a wavelength of 3 mm) in configuration C40-3, which included 38$\times$12m antennae and 8$\times$7m antennae. However, 3$\times$12m and 1$\times$7m antennae failed in calibration, leaving 42 antennae for imaging. A single quiet-Sun target about 200\arcsec\ south-west of disk center, tracked for solar rotation, was observed in 10.5 minute scans separated by 2-minute calibration scans, with 2-second integrations, for a total of 37 minutes. The Sun was observed in the time interval 16:00--16:45 UT (45 minutes). Interferometric data were supplemented by single-dish images of the full Sun with a cadence of $\approx$ 10 minutes, made with ALMA's total power antennae at a resolution of 60\arcsec. Individual 2-second interferometric images were mapped and self-calibrated using standard CASA software, then restored with a synthesized beam of 1.6\arcsec\ and corrected for the primary beam response. The four 2 GHz-wide spectral windows at 93, 95, 105 and 107 GHz were combined in the mapping, as the resulting $u,v$-coverage was found to improve the images. The mapped field of view (FOV) was about 120\arcsec, in order to accomodate the fact that the 7m dishes have a primary beam (full-width at half-maximum, or FWHM) of 100\arcsec, while the 12m-dishes have a FWHM of 58\arcsec\ at 100 GHz. The images were made 401 pixels square, with a cell size of 0.3\arcsec\ and a cadence of 2 seconds. The final data cube $T_b(x,y,t)$ contains 1090 images. Solar observing with ALMA is described in the commissioning papers by \citet{2017SoPh..292...87S} and \citet{2017SoPh..292...88W}, and calibrations followed the prescriptions given therein: in particular, the interferometric images were converted from flux to brightness temperature, and then 7236 K (derived from the target location in the single-dish images) was added to each image to account for the fact that the interferometer resolves out large spatial scales. 
%\subsection{Other data and data alignment} 
 
Co-temporal data at other wavelengths used in this study comprised 60 photospheric magnetograms with a 45-sec cadence from the Helioseismic and Magnetic Imager \citep[HMI, ][]{2012SoPh..275..207S} on board the SDO, 220 images at 1600 \AA, 1700 \AA, and 304 \AA,  obtained with a 12-sec cadence by the Atmospheric Imaging Assembly \citep[AIA, ][]{2012SoPh..275...17L} on board the SDO, and 82 IRIS slit-jaw images with a 32-sec cadence in the 2796 \AA\ Mg II k line. The data series was completed with full-disk images in $H\alpha$: 38 images from the BBSO and 35 images from the GONG telescope at CTIO, both taken with a 0.5\AA\--0.6\AA\ bandpass at a 1-min cadence and a nominal resolution of 2\arcsec.  
 
The full disk SDO images were chosen as the coordinate reference against which all other images were coaligned by cross-correlating images averaged over 45 minutes. The remaining image displacements are less than 1\arcsec\ and do not affect the results of the analysis presented here.  The data from the different instruments were analysed at their original spatial resolution and in their original data format (data numbers, or DN), except for the ALMA data, which was translated into absolute brightness or temperature values (Kelvin) as described above.

\begin{figure*} 
  \centering 
            \includegraphics[width=0.6\textwidth]{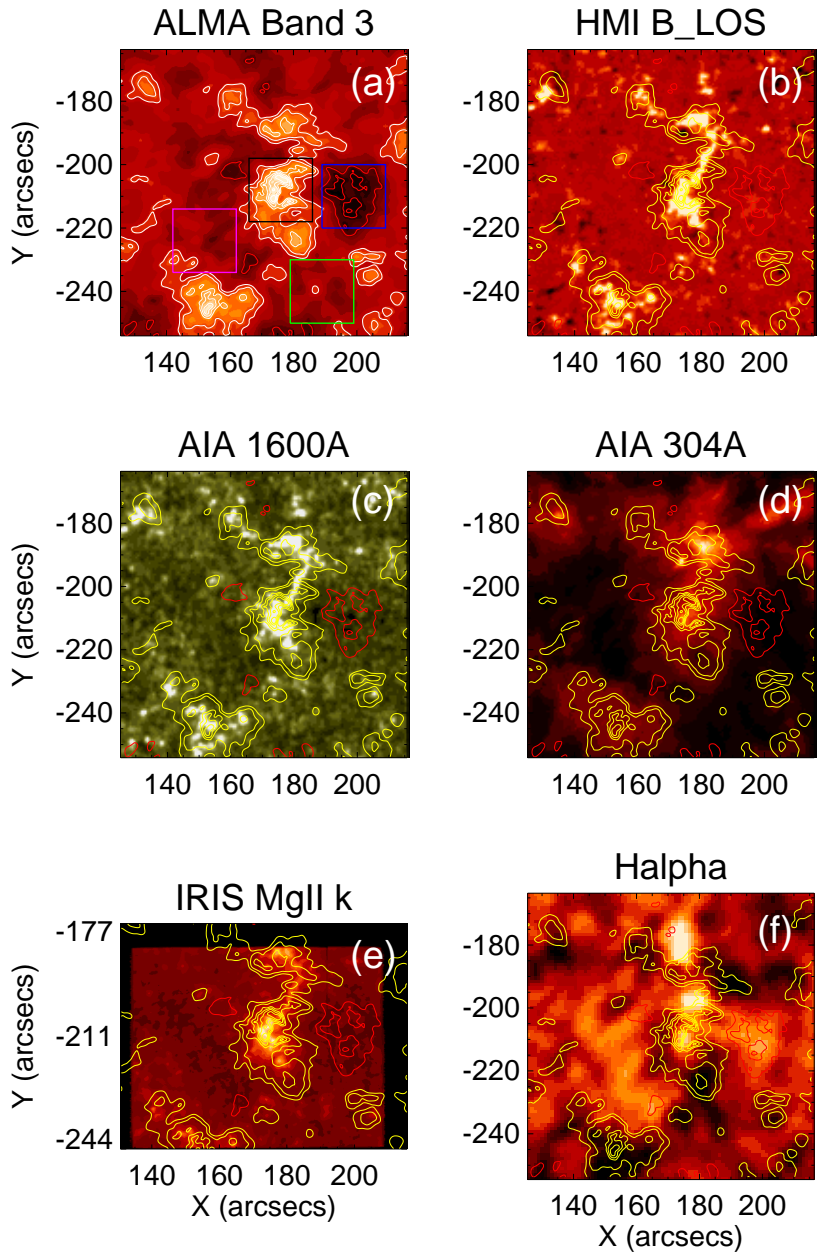}%{fig_maps_mean_all_ha_sh1600_2_with_subregions.eps} 
         \caption{The analyzed quiet-Sun region: (a) ALMA Band 3 ($\lambda \approx\,$3~mm) image, (b) SDO/HMI magnetogram saturated outside the range [-50,50]G;  (c) and (d) SDO/AIA images in the 1600 \AA\ and He II 304 \AA\ channels, (e) IRIS Mg II k 2796 \AA\ image, and (f) CTIO $H\alpha$ image. The AIA images are clipped at 50\% of maximum brightness. The overlaid color contours indicate ALMA brightness temperatures at  7500, 7800, 8300, 8500, 8700, 8900~K (yellow), and at 6000 and 6500~K (red). All images are averaged over 45 minutes.  Color rectangles in panel (a) indicate the subregions of interest: a dark region D (blue), a bright network region NW (black), and two internetwork regions, IN1 (magenta) and IN2 (green). Display range of the ALMA image is from 5500 to 9000~K. 
              } 
         \label{fig_1} 
   \end{figure*}

\section{Results}\label{res} 
 
\subsection{Time-averaged images} 
 
For morphological comparison of the chromospheric features seen in different spectral domains we utilized mean images, obtained by averaging in time over the whole observing period. In this way we avoid biasing our results due to very short-lived features. Figure~\ref{fig_1} depicts the time-averaged quiet solar chromosphere at different heights from the temperature minimum region (SDO/AIA 1600 \AA) through the middle chromosphere (IRIS 2796 \AA\ Mg II k line and ALMA Band 3) to the upper chromosphere (CTIO $H\alpha$) and the transition region (SDO/AIA 304 \AA),  as well as the corresponding SDO/HMI line-of-sight photospheric magnetogram. In the time-averaged Band 3 image the brightness range is from 5630~K to 9140~K. The time-averaged magnetogram values are within the range of  [-70, 540]~G.  
 
\begin{figure*} 
  \centering 
            \includegraphics[width=0.65\textwidth]{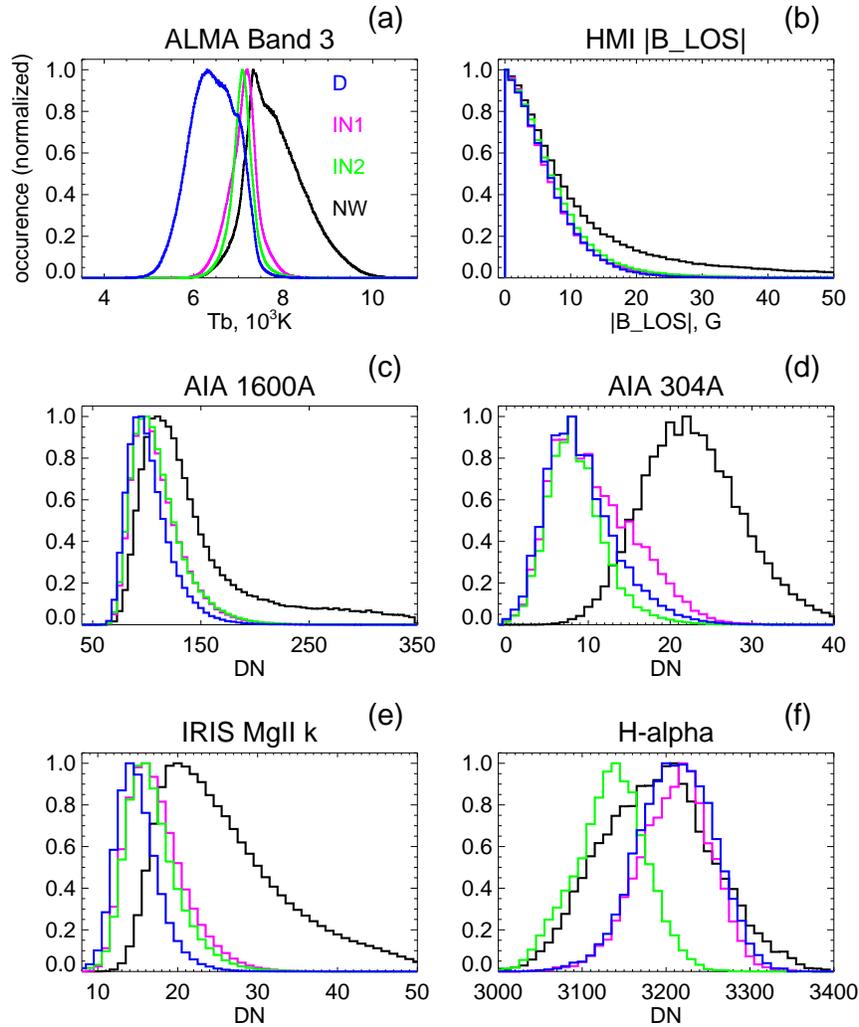}%{histo_sudregions_6_hac_sh1600_2_upd.eps} 
      \caption{Intensity histograms for {\textbf the pixels in} the four coloured squares in Fig.~\ref{fig_1}, with each histogram covering the full cube of (a) ALMA Band 3 (3~mm) data, (b) SDO/HMI magnetograms,  (c) and (d) SDO/AIA  images in the 1600 \AA\ and 304 \AA\ channels, (e) IRIS 2796 \AA\ Mg II k  images, and (f) CTIO $H\alpha$ images. The color code corresponds to  Fig.~\ref{fig_1} .
              } 
         \label{fig_2} 
\end{figure*}

\begin{figure*} 
  \centering 
            \includegraphics[width=0.6\textwidth]{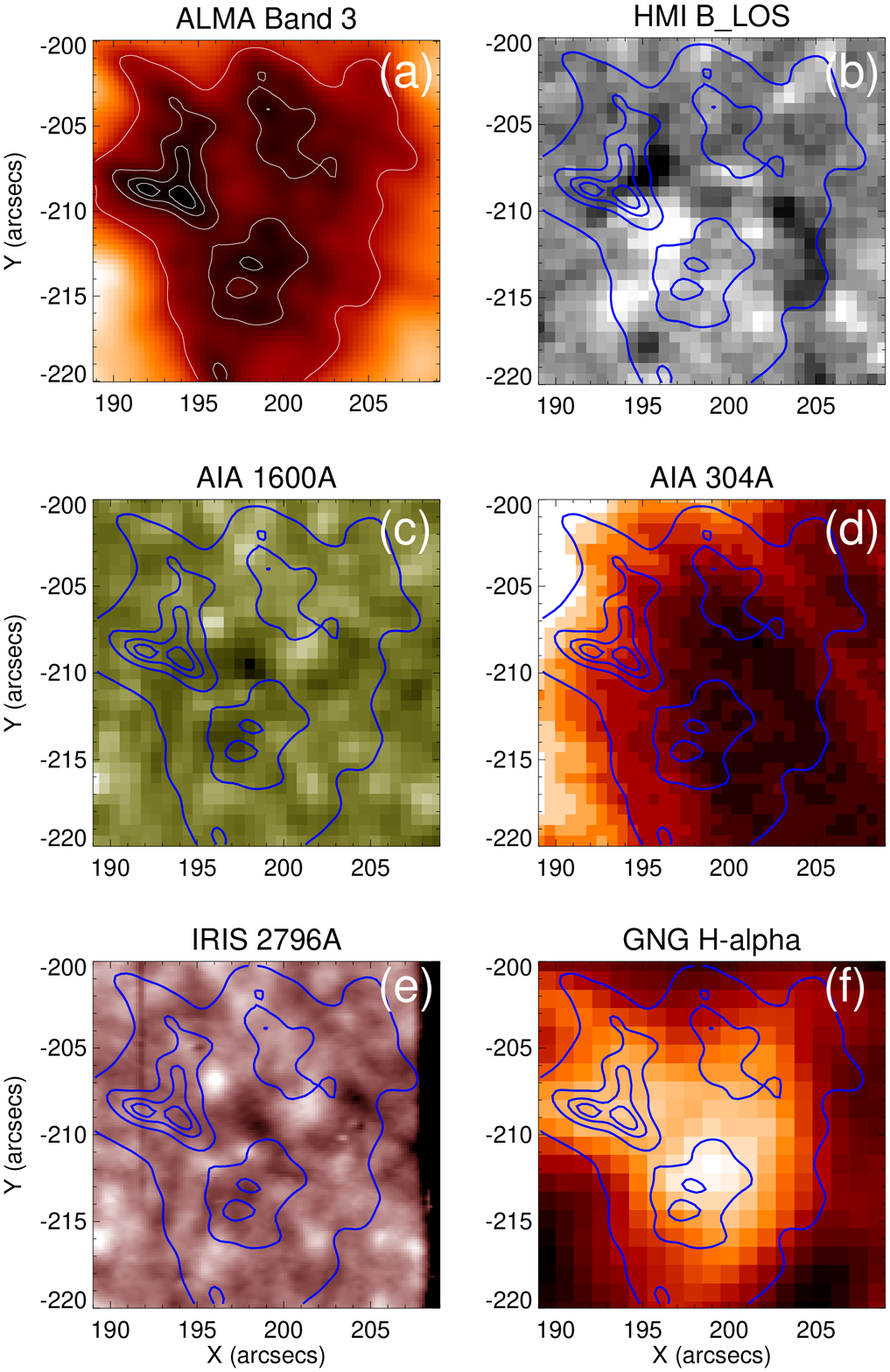}%{b3_sdo_iris_ha_DARK_sh1600_2_hac.eps} 
      \caption{Time-averaged ChAH images at 6 wavelengths:  (a) ALMA Band 3, (b) photospheric magnetogram saturated outside the range [-10,10]G, (c) and (d) AIA channels at 1600 \AA\ and 304 \AA, respectively,  (e) Mg II k 2796 \AA, and (f) $H\alpha$. The overlaid color contours indicate ALMA brightness temperatures at 5700, 5800, 6000, and 6500~K. The plotted FOV corresponds to the blue rectangle in Fig.~\ref{fig_1}. 
              } 
         \label{fig_3} 
\end{figure*}

Bright network is the most prominent feature in all time-averaged images in Fig.~\ref{fig_1}. Enhanced Band 3 emission, shaped like a ``seahorse'' formed by the six brightness contours between 7500~K and 8900~K in Fig.~\ref{fig_1}, coincides with enhanced magnetic flux and increased brightness in other panels, and outlines  bright chromospheric network patches. Striking is the presence of a particularly dark (cool) area to the right of the ALMA image center (outlined by red contours at 6000~K and 6500~K, and located within a blue rectangle in Fig.~\ref{fig_1}a)  that is  not distinguishable in the UV channels or in the magnetogram.   
Hereafter we refer to this dark area as a CHromospheric ALMA Hole (ChAH), since it is apparent only in ALMA images. 
To study the ChAH phenomenon, we complemented it with three other regions of the same size in the FOV, representing chromospheric network (black rectangle in Fig.~\ref{fig_1}a) and two regions of  ``average brightness'' corresponding to ``normal internetwork'' (green and magenta rectangles in Fig.~\ref{fig_1}a), and compared observational data for these four regions of interest at different wavelengths. The results of this comparison for data sequences (in the form of histograms) and for images integrated over the full sequence duration are shown in Figs.~\ref{fig_2}-\ref{fig_3},  and are discussed in the following subsections. 
 
\subsection{Time sequences of images} 
 
In the full ALMA data cube $T_b(x,y,t)$ the brightness range is from 4370~K to 11170~K with the coolest gas being found within the ChAH. Locations in the ChAH region exhibit localized brightness changes with time, but typically only over a range of about 1000 K. In 57\% of the 2s frames there are regions within the blue ChAH rectangle where the brightness temperature is below 5000~K. In general the ChAH remains quite stable over the whole time series, demonstrating that it is a long lasting phenomenon not related to dynamic effects. %The signal in the full-FOV magnetogram sequence lies within the range of [-130, 630]~G.} 
 
To confirm the ChAH's stability over the duration of the observations and to study whether the ChAH differs from the other selected regions of interest at any of the considered wavelengths, we constructed brightness and data numbers histograms. The histograms  built from the full time sequence of images at 6 wavelengths are shown in  Fig.~\ref{fig_2}. The dark ChAH region, marked as ``D'', is plotted in blue,  normal internetwork (``IN1'' and ``IN2'') are in magenta and   in green, respectively, and the chromospheric network patch (``NW'') is in black.  
 
The distribution of ALMA brightness (Fig.~\ref{fig_2}a) is significantly different from the other wavelengths. The SDO and IRIS data cubes used here are characterized by distributions with a pronounced high-intensity tail, resembling a log-normal, which is expected for histograms of EUV and UV intensities with sufficiently good statistics  \citep{2000A&A...362..737P}. Band 3 brightness histograms, on the contrary, are largely symmetric.  Similar symmetric forms of the ALMA brightness at a shorter wavelength of 1.3~mm were found by \citet{2019arXiv190105763J}, while a histogram with a pronounced high-temperature tail was obtained in \citet{2009A&A...497..273L} from quiet-Sun BIMA brightness at 3 mm. 
 
Some asymmetry can be seen in the Band 3 histogram for the NW region (black curve in Fig.~\ref{fig_2}a), which has more pixels in the high-temperature part,  mainly due to magnetic features. Surprisingly,  regions with ``average'' brightness, IN1 and IN2, show narrow and symmetric distributions, with no clear signatures of shocks.  
The ChAH region (D) has a significantly wider histogram in Band 3 with the histogram maximum located about 1000~K below those of the histograms of the other regions.  Striking is that this significant difference between the ChAH and internetwork intensity distributions is seen only at the mm wavelength; in the other spectral channels the ChAH histograms are similar to those of the IN1 and IN2 regions, except in the magnesium line (Fig.~\ref{fig_2}e). There the ChAH histogram is slightly narrower due to the lack of bright points as compared to IN1 and IN2, but is not clearly offset as in the ALMA data. In $H\alpha$, the intensity distributions in D, IN1, and NW are similar to a large extent. Consequently, the ChAH, discovered in the ALMA Band 3 data, is indistinguishable in the full-cube intensity histograms of all the other wavelengths considered.   
 
\subsection{Chromospheric ALMA Hole}\label{comp} 
 
Figure~\ref{fig_3} depicts time-averaged ChAH images at six wavelengths. With horizontal dimensions of approx. 20\arcsec\ x 20\arcsec\ the ChAH is  intermediate in size, between the meso- and supergranule size scale. The brightness temperature ranges from 5600~K to 7500~K in the Band 3 image shown in Fig~\ref{fig_3}a, while in the ChAH data cube it drops as low as 4370~K, which is about 60\% of the quiet-Sun brightness temperature at this wavelength \citep{2017SoPh..292...88W}. The magnetogram signal in the area is weak, staying within the [-110, 78]~G range in the full time sequence. According to Fig.~\ref{fig_2}b the ChAH is associated with a similar amount of magnetic field as found in normal internetwork areas.  Magnetic concentrations do not appear to avoid the ChAH (Fig. 3b), unlike the interiors of mesogranular structures \citep{2011ApJ...727L..30Y}. Nor does it seem to be related to the ``dead calm'' magnetic areas studied by \citet{2012ApJ...755..175M} and characterized by the near absence of magnetic bipoles. As can be seen from the overlays of mm contours in Fig.~\ref{fig_3} there is no one-to-one correspondence between the localized depressions in Band 3 brightness and the distribution of flux in the UV images. The AIA 304~\AA\ channel also displays a low brightness in this region (Fig.~\ref{fig_3}d), although not lower than in other IN regions (Fig.~\ref{fig_2}d), while in $H\alpha$  the ChAH region is relatively bright (Fig.~\ref{fig_3}f).

\section{Discussion and Conclusions}\label{discussion} 
 
ALMA data not only distinguish clearly between network and internetwork, they also show regions that are particularly dark, which we call Chromospheric ALMA Holes (ChAH).  Comparing the Band 3 data with those in 7 wavelengths, covering the full range of chromospheric heights from the temperature minimum to the bottom of the transition region, we conclude that no other data set shows even nearly as clear a difference between ChAH and normal internetwork as ALMA does. In the Mg II k-line, 304 \AA, 1600 \AA, and 1700~\AA\ the ChAH region looks equally as dark as normal IN.  In the ALMA data, on the contrary,  the minimum intensity in the ChAH region (4370~K) is well below the values in other pixels. $H\alpha$ images look significantly different than Band 3 or any of the other data considered in this work. The fact that the ChAH region is bright in $H\alpha$ implies that $H\alpha$ and Band 3 sense different parts of the chromosphere. 
 
In this letter we demonstrated that ALMA detects chromospheric features missed by the SDO and IRIS. For the SDO this is not surprising, as  in the quiet Sun the HMI line Fe I 6173 \AA,  and the AIA 1600 \AA, and 1700 \AA\ are mainly formed at heights considerably below ALMA Band 3, while 304~\AA\ is formed considerably above, so that the SDO samples different heights (and temperatures) than ALMA's Band 3 does. Also, the Lyman continuum absorbs any photons emitted by He II 304 \AA\ in the chromosphere, so that this line is insensitive to this layer. Similarly, H$\alpha$ typically forms high in the chromosphere, and responds to any hot gas present there, but is not expected to have much contribution from cool gas because of lack of excitation of its ground level, $n=2$. Mg II k-line images are more relevant for a comparison with the Band 3 data, as the radiation at these wavelengths is believed to be formed over similar ranges of heights \citep[e.g., ][]{2018A&A...620A.124D} and the strength of the central reversal of the line core is  thought to follow the temperature quite closely \citep[although extremely nonlinearly, see, e.g., ][]{2013ApJ...772...90L}. However, because the core of this line follows temperature exponentially (in the Wien limit), it is expected to be extremely insensitive to cool gas, such as found in the ChAH. The unexpectedly symmetrical form of the Band 3 brightness histograms, in contrast to the roughly log-normal shape of the UV radiation histograms, can be due to the linear relation between gas temperature and brightness at mm wavelengths, in contrast to the highly non-linear relationship in the UV noted above. 
 
It appears that ALMA, through its very different sensitivity to temperature, has sensed cool chromospheric gas, which has escaped other diagnostics at shorter wavelengths. Thus, the ALMA observations presented here have confirmed that the mm continuum is a unique thermal diagnostic for the middle chromosphere. The detection of such low temperatures in ALMA images demonstrates ALMA's potential to differentiate between coolish and really cool parts of the chromosphere, which the UV lines cannot do.  In UV observations, scattered light is a significant problem and the excess radiation from some parts of the chromosphere can leak into darker regions due to the huge brightness contrasts. Also, the UV lines pick up the brightest plasma along the LOS and may not be able access the coolest features \citep[e.g., ][]{1995ApJ...440L..29C}. The presence of very cool gas has been proposed by, e.g., \citet{1981ApJ...244.1064A}, \citet{1986ApJ...304..542A}, \citet{1994Sci...263...64S}, and \citet{2002ApJ...575.1104A} based on observations of fundamental-band vibration-rotation transitions of CO in the infrared. The ALMA observations return higher minimum temperatures than inferred from CO lines. This may have two reasons. Firstly, ALMA Band 3 is formed in the middle chromosphere \citep{2004A&A...419..747L, 2017A&A...601A..43L}, while the CO lines mainly sample deeper layers \citep[e.g., ][]{2014LRSP...11....2P}. Secondly, it has to do with the fact that ALMA reacts linearly to temperature, while the CO fundamental band lines increase non-linearly in strength with decreasing temperature. Therefore, we expect in an inhomogeneous atmosphere with hotter and cooler gas along the LOS, the ALMA observations to find temperatures in between those found from CO lines and atomic UV lines. A study of the quiet Sun similar to the present one, but including ALMA Bands 6 and 7, which sample deeper layers of the chromosphere, and CO lines would be of considerable value.  
 
\acknowledgments 
 
This paper makes use of the following ALMA data: $ADS/JAO.ALMA\#2016.1.00202.S$. ALMA is a partnership of ESO (representing its member states), NSF (USA) and NINS (Japan), together with NRC (Canada) and NSC and ASIAA (Taiwan) and KASI (Republic of Korea), in cooperation with the Republic of Chile. The Joint ALMA Observatory is operated by ESO, AUI/NRAO and NAOJ.  The National Radio Astronomy Observatory is a facility of the National Science Foundation operated under cooperative agreement by Associated Universities, Inc. 
IRIS is a NASA Small Explorer mission developed and operated by LMSAL with mission operations executed at NASA Ames Research center and major contributions to downlink communications funded by ESA and the Norwegian Space Centre. The AIA and HMI data are courtesy of the NASA/SDO, as well as AIA and HMI science teams. This work utilizes data obtained by the Global Oscillation Network Group (GONG) Program, managed by the National Solar Observatory, which is operated by AURA, Inc. under a cooperative agreement with the National Science Foundation. The data were acquired by instruments operated by the Big Bear Solar Observatory and Cerro Tololo Interamerican Observatory. 
The work was performed within the SAO RAS state assignment in the part ''Conducting Fundamental Science Research''. Part of the work has been supported by Russian RFBR grant 18-29-21016. This work has been partially supported by the BK21 plus program through the National Research Foundation (NRF) funded by the Ministry of Education of Korea. 
 
\bibliographystyle{apj} 
\bibliography{loukitcheva} 
\end{document}